\documentclass[aps,pra,twocolumn,groupedaddress]{revtex4-1}
\usepackage{amsmath,amssymb}
\usepackage{bm}
\usepackage{ulem}
\usepackage{xcolor}
\usepackage{graphicx}
\usepackage[dvipdfmx]{hyperref}
\hypersetup{colorlinks=true,citecolor=purple,linkcolor=magenta,urlcolor=purple}
\begin{document}

% Use the \preprint command to place your local institutional report
% number in the upper righthand corner of the title page in preprint mode.
% Multiple \preprint commands are allowed.
% Use the 'preprintnumbers' class option to override journal defaults
% to display numbers if necessary
%\preprint{}

%Title of paper
\title{Suppressing nonadiabatic transitions during adiabatic generation of highly entangled states in bosonic Josephson junctions}

% repeat the \author .. \affiliation  etc. as needed
% \email, \thanks, \homepage, \altaffiliation all apply to the current
% author. Explanatory text should go in the []'s, actual e-mail
% address or url should go in the {}'s for \email and \homepage.
% Please use the appropriate macro foreach each type of information

% \affiliation command applies to all authors since the last
% \affiliation command. The \affiliation command should follow the
% other information
% \affiliation can be followed by \email, \homepage, \thanks as well.
\author{Takuya Hatomura}
\email[]{takuya.hatomura.ub@hco.ntt.co.jp}
%\homepage[]{Your web page}
%\thanks{}
%\altaffiliation{}
\affiliation{NTT Basic Research Laboratories, NTT Corporation, Kanagawa 243-0198, Japan}

%Collaboration name if desired (requires use of superscriptaddress
%option in \documentclass). \noaffiliation is required (may also be
%used with the \author command).
%\collaboration can be followed by \email, \homepage, \thanks as well.
%\collaboration{}
%\noaffiliation

\date{\today}

\begin{abstract}
We study suppression of nonadiabatic transitions during adiabatic generation of a cat-like state (a superposition of different size cat states) and a spin squeezed state in a bosonic Josephson junction. 
In order to minimize the adiabatic error, we use quantum adiabatic brachistochrone, which enables us to track a geometrically efficient path in parameter space under given conditions without requiring additional terms. 
For creation of a cat-like state, divergence of the quantum geometric tensor associated with gap closing at the critical point is avoided because of the parity conservation. 
The resulting schedules of parameters are smooth and monotonically decreasing curves. 
Use of these schedules offers reduction of time to generate both of a cat-like state and a spin squeezed state. 
\end{abstract}

% insert suggested PACS numbers in braces on next line
% insert suggested keywords - APS authors don't need to do this
%\keywords{}

%\maketitle must follow title, authors, abstract, \pacs, and \keywords
\maketitle

% body of paper here - Use proper section commands
% References should be done using the \cite, \ref, and \label commands

%
%==================================================================
%
\section{Introduction}

Development of quantum science and technologies has elucidated usefulness of entanglement. 
In many fields, use of entangled states enables us to outperform classical devices~\cite{Acin2018}. 
For example, in quantum metrology, we can produce sensors that estimate unknown parameters with higher resolution than the classical limitation called the standard quantum limit~\cite{Giovannetti2011,Degen2017,Pezze2018}. 
Moreover, macroscopic entanglement provides us with significant resolution approaching the quantum limitation known as the Heisenberg limit~\cite{Giovannetti2011,Degen2017,Pezze2018}. 
Unknown parameters that we can estimate differ depending on physical systems, and thus generation of highly entangled states has been investigated in various systems. 
In atomic systems, a lot of successful experiments generating spin squeezed states have been reported~\cite{Esteve2008,Gross2010,Riedel2010,Maussang2010,Ockeloen2013,Strobel2014,Muessel2014,Muessel2015,Schmied2016}. 
The number of particles composing these spin squeezed states is suitable to be said macroscopic (see, e.g., Ref.~\cite{Pezze2018}). 
Moreover, cat states have been also produced in trapped ions~\cite{Leibfried2005,Monz2011,Omran2019}. 
However, in contrast to spin squeezed states, the number of particles composing cat states is still limited to a dozen.

Typically, these highly entangled states can be dynamically generated by using the one-axis twisting interaction~\cite{Kitagawa1993}. 
Indeed, the one-axis twisting interaction leads a coherent spin state into a spin squeezed state at first~\cite{Kitagawa1993} and into a cat state later~\cite{Molmer1999}. 
However, generation time of a cat state is not so short, and thus it is difficult to create a large cat state within coherence time. 
Faster generation can be realized by just adding the Rabi coupling to the one-axis twisting interaction~\cite{Micheli2003}, while fidelity to a cat state decreases when system size becomes large. 
After these generation schemes were proposed, some tricks to improve dynamical generation have been discussed~\cite{Carr2010,Garcia-March2011}.
Not only generation time and size of entanglement, but also to stop creation at final time is also difficult. 
Because these generation schemes are based on dynamical interference driven by nonlinear interactions, we have to turn off all interactions at final time. 
Otherwise, highly entangled states collapse into useless states.

These highly entangled state can be also generated by adiabatic driving. 
Indeed, with certain parameter values, the ground state of a system consisting of the one-axis twisting interaction and the Rabi coupling is a spin squeezed state for repulsive interaction~\cite{Steel1998} and that is a cat state for attractive interaction~\cite{Cirac1998}. 
Note that it is difficult to directly approach these ground states by just cooling down because of small energy gaps. 
In adiabatic generation, we first prepare the trivial ground state in the large Rabi coupling limit, and then we adiabatically sweep parameters into the large one-axis twisting interaction limit~\cite{Cirac1998,Leggett1998,Javanainen1999,Lee2006,Yukawa2018}. 
Notably generation time of a cat state is comparable to typical coherence time~\cite{Yukawa2018}. 
However, the size of generated cat states is still up to $N=20$~\cite{Omran2019}. 
This limitation mainly arises from particle losses~\cite{Spehner2014,Pawowski2017}. 
Particle losses induce various noises during generation and destroy entanglement. 
Because particle losses are inevitable in ultracold atom experiments, speedup of generation is rather of interest to reduce bad influence.

In both of dynamical and adiabatic generation of spin squeezed states, application of optimal control theory and shortcuts to adiabaticity has been discussed and speedup of generation was found~\cite{Grond2009a,Julia-Diaz2012,Yuste2013,Pichler2016,Stefanatos2018}. 
However, optimal control theory usually requires fast and irregular oscillating schedules of parameters~\cite{Grond2009a,Pichler2016}, which induce nonadiabatic transitions and interference to finally achieve high fidelity to a target state, and thus experimental implementation is not so straightforward. 
Shortcuts to adiabaticity also requires oscillating schedules of parameters, whereas oscillation of these schedules is relatively slow~\cite{Julia-Diaz2012,Yuste2013,Stefanatos2018}. 
For creation of cat states, application of optimal control theory and shortcuts to adiabaticity was also considered~\cite{Lapert2012,Hatomura2018a}. 
In the former case, fast and irregular oscillating schedules and turning off all interactions at last are required to accelerate dynamical generation~\cite{Lapert2012}. 
In latter case, the two-axis countertwisting interaction~\cite{Kitagawa1993} is additionally required to speedup adiabatic generation~\cite{Hatomura2018a}. 
However, in spite of the fact that usefulness of the two-axis countertwisting interaction for fast generation of macroscopic entanglement was pointed out about a quarter century ago~\cite{Kitagawa1993} and a lot of methods for realization were proposed~\cite{Liu2011,Shen2013,Zhang2014,Opatrny2015,Kajtoch2016}, there is no experimental realization of it in atomic systems yet. 
Therefore, other realistic routes to fast generation should be investigated. 
It should be also worthy to mention that in the context of shortcuts to adiabaticity generation of this type of a cat state was first discussed in the transverse Ising model, where not only two-axis countertwisiting-like interaction but also much higher order interactions are required~\cite{DelCampo2012}.

Quantum adiabatic brachistochrone is a systematic method to find a time-optimal schedule of parameters in a system~\cite{Rezakhani2009,Rezakhani2010}. 
There, a cost function of the adiabatic condition is viewed as the action and minimized according to the variational principle. 
The resulting Euler-Lagrange equation gives the geodesic equation for adiabaticity and provides a schedule of parameters. 
Compared with counterdiabatic driving~\cite{Demirplak2003,Berry2009}, which is one of the methods in shortcuts to adiabaticity~\cite{Guery-Odelin2019}, quantum adiabatic brachistochrone does not require additional terms, and thus it is rather implementable.

In this paper, we use quantum adiabatic brachistochrone to suppress nonadiabatic transitions during adiabatic generation of a cat-like state and a spin squeezed state in a bosonic Josephson junction without requiring additional terms. 
We show that both optimized schedules for adiabatic generation of a cat-like state and a spin squeezed state are smooth and monotonically decreasing curves. 
Here, for creation of a cat-like state, we use the parity conservation of a bosonic Josephson junction to avoid divergence of the quantum geometric tensor. 
According to these schedules, we observe increase of quantum Fisher information for a generated cat-like state and decrease of the spin squeezing parameter for a generated spin squeezed state, which represent improvement of adiabaticity and increase of metrological usefulness.

This paper is constructed as follows. 
In Sec.~\ref{Sec.method}, we review properties of bosonic Josephson junctions and formalism of quantum adiabatic brachistochrone. 
Generation schemes are explained in Sec.~\ref{Sec.setup} in detail. 
We first discuss generation of cat-like states in Sec.~\ref{Sec.cat} and later discuss generation of spin squeezed states in Sec.~\ref{Sec.squeezed}. 
We summarize the present article in Sec.~\ref{Sec.summary}.

%
%==================================================================
%
\section{\label{Sec.method}Methods}
%
%-----------------------------------------------------------------
%
\subsection{Bosonic Josephson junctions}
Bosonic Josephson junctions consist of two species of bosons $a_1$ and $a_2$, which realize in various systems including the low-energy limit of  Bose-Einstein condensates~\cite{Gati2007}. 
The Hamiltonian of a bosonic Josephson junction is given by
\begin{equation}
\mathcal{H}_\mathrm{BJJ}=\hbar\chi J_z^2+\hbar\Omega J_x,
\label{Eq.BJJ.ham}
\end{equation}
where $\chi$ is the Kerr non-linearity, $\Omega$ is the Rabi coupling, and $J_\alpha$, $\alpha=x,y,z$ is the angular momentum representation of bosonic operators
\begin{equation}
\left\{
\begin{aligned}
&J_x=\frac{1}{2}(a_1^\dag a_2+a_2^\dag a_1), \\
&J_y=\frac{1}{2i}(a_1^\dag a_2-a_2^\dag a_1), \\
&J_z=\frac{1}{2}(a_1^\dag a_1-a_2^\dag a_2). 
\end{aligned}
\right.
\end{equation}
The first term is also known as the one-axis twisting interaction~\cite{Kitagawa1993}. 
This system is characterized by the parameter
\begin{equation}
\Lambda=\frac{\chi N}{\Omega},
\end{equation}
where $N$ is the number of atoms. 
Here we assume $\Omega\le0$.

For positive nonlinearity $\chi>0$, the ground state of the Hamiltonian (\ref{Eq.BJJ.ham}) becomes a spin squeezed state, where fluctuation along the $z$-axis, $\Delta J_z$, is suppressed and fluctuation along the $y$-axis, $\Delta J_y$, is enhanced, when the parameter $\Lambda^{-1}$ is small, $|\Lambda^{-1}|<1$~\cite{Steel1998}. 
This type of spin squeezed states has been realized in experiments by adiabatically splitting condensates~\cite{Esteve2008}. 
The level of squeezing depends on competition between adiabaticity and losses, i.e., how slowly changing parameters to suppress nonadiabatic transitions and how quickly generating a spin squeezed state to suppress bad influence of particle losses~\cite{Maussang2010}. 
Therefore, speedup of generation is of interest.

In contrast, for negative nonlinearity $\chi<0$, a cat state, a superposition of a mode-1 condensate and a mode-2 condensate, can be realized as the ground state of the Hamiltonian (\ref{Eq.BJJ.ham}) for $0\le\Lambda^{-1}<1$~\cite{Cirac1998}. 
However, creation of this cat state would be failed if one just cools the system in this parameter region due to occurrence of spontaneous symmetry breaking. 
This happens because of the exponentially small energy gap between the ground state and the first excited state. 
Adiabatic generation is one of the strategies to create this cat state~\cite{Cirac1998,Lee2006,Yukawa2018}. 
The key point of adiabatic generation is the parity conservation~\cite{Yukawa2018} ensured by the commutation relation
\begin{equation}
[\mathcal{H}_\mathrm{BJJ},\Pi]=0,
\end{equation}
where $\Pi$ is the parity operator
\begin{equation}
\Pi=\exp[i\pi(J-J_x)]. 
\end{equation}
In adiabatic generation, we first start with the trivial ground state of the Hamiltonian (\ref{Eq.BJJ.ham}) in the disordered phase $\Lambda^{-1}>1$ that has the parity $\Pi=+1$. 
Because the first excited state has the parity $\Pi=-1$ and the parity is conserved in time evolution, we can ignore the exponentially small energy gap~\cite{Yukawa2018}. 
The parity conservation also ensures that the generated state is rather a superposition of different size cat states~\cite{Hatomura2019},
\begin{equation}
\begin{aligned}
&|\Psi\rangle=\sum_mg_m|\Psi_m\rangle, \\
&|\Psi_m\rangle=\frac{1}{\sqrt{2}}(|N-m,m\rangle+|m,N-m\rangle),
\end{aligned}
\end{equation}
where $\sum_m|g_m|^2=1$. 
Therefore, the generated state is close to a cat state (the NOON state) if $\{|g_m|^2\}$ is distributed around $m\approx0$. 
It was shown that the parity measurement and the Fourier analysis enable us to detect this superposition including the information of the size distribution of these cat states $\{|g_m|^2\}$ and also to maximally extract the potential of this superposition in interferometry~\cite{Hatomura2019}. 
However, these properties are lost in an exponential way under particle losses although its decay rate is smaller than that of dynamical generation~\cite{Hatomura2019}, and thus we are interested in acceleration of generation.

%
%-----------------------------------------------------------------------
%
\subsection{Quantum adiabatic brachistochrone}
Quantum adiabatic brachistochrone provides us with systematic temporal optimization of Hamiltonians to achieve adiabatic time evolution~\cite{Rezakhani2009,Rezakhani2010}. 
Here we briefly review this method.

For a given Hamiltonian $\mathcal{H}(\bm{\lambda};s)$, the dynamical transformation is a unitary operator $V(s)$ that defines time evolution of a given state
\begin{equation}
|\Psi(s)\rangle=V(s)|\Psi(0)\rangle,
\end{equation}
by the Schr\"odinger dynamics
\begin{equation}
i\hbar\frac{\partial}{\partial s}V(s)=T\mathcal{H}(\bm{\lambda};s)V(s),
\end{equation}
where $s=t/T$ is the normalized time and $T$ is the operation time. 
Here we assume that the Hamiltonian depends on time through time-dependent parameters $\bm{\lambda}(s)=(\lambda_1,\lambda_2,\cdots;s)$. 
For the same Hamiltonian, the adiabatic transformation is a unitary operator $V_\mathrm{ad}(s)$ that isometrically transforms the projection operators as
\begin{equation}
V_\mathrm{ad}(s)P_n(\bm{\lambda};0)V_\mathrm{ad}^\dag(s)=P_n(\bm{\lambda};s),
\end{equation}
for all $n$, where
\begin{equation}
\mathcal{H}(\bm{\lambda};s)P_n(\bm{\lambda};s)=E_n(\bm{\lambda};s)P_n(\bm{\lambda};s),
\end{equation}
with the eigenenergy $E_n(\bm{\lambda};s)$. 
The adiabatic transformation generates adiabatic time evolution
\begin{equation}
|\Psi_\mathrm{ad}(s)\rangle=V_\mathrm{ad}(s)|\Psi(0)\rangle. 
\label{Eq.ATE}
\end{equation}
There exists a Hamiltonian $\mathcal{H}_\mathrm{ad}(s)$ that generates adiabatic time evolution (\ref{Eq.ATE}) as the Schr\"odinger dynamics
\begin{equation}
i\hbar\frac{\partial}{\partial s}V_\mathrm{ad}(s)=T\mathcal{H}_\mathrm{ad}(s)V_\mathrm{ad}(s).
\end{equation}
This adiabatic Hamiltonian $\mathcal{H}_\mathrm{ad}(s)$ is given by
\begin{equation}
\mathcal{H}_\mathrm{ad}(s)=\mathcal{H}(\bm{\lambda};s)+\frac{i\hbar}{T}[\partial_{s}P_n(\bm{\lambda};s),P_n(\bm{\lambda};s)],
\label{Eq.ad.ham.n}
\end{equation}
for a given state in the $n$th eigensector of the Hamiltonian $\mathcal{H}(\bm{\lambda};s)$~\cite{Avron1987}. 
Note that, in the context of shortcuts to adiabaticity~\cite{Guery-Odelin2019}, the additional term is nothing but the single-spectrum counterdiabatic term~\cite{Takahashi2013}. 
If we consider the adiabatic transformation of the full eigensectors of the Hamiltonian, we should introduce the usual counterdiabatic terms~\cite{Demirplak2003,Berry2009} instead of Eq.~(\ref{Eq.ad.ham.n}).

Deviation of the dynamical transformation from the adiabatic transformation can be estimated by using the wave operator
\begin{equation}
\Omega(s)=V_\mathrm{ad}^\dag(s)V(s), 
\end{equation}
satisfying the Volterra equation
\begin{equation}
\Omega(s)=1-\int_0^sK_T(s^\prime)\Omega(s^\prime)ds^\prime, 
\end{equation}
with the kernel
\begin{equation}
K_T(s)=V_\mathrm{ad}^\dag(s)[\partial_{s}P(\bm{\lambda};s),P(\bm{\lambda};s)]V_\mathrm{ad}(s), 
\end{equation}
where $\Omega(s)$ is close to $1$ when the dynamical transformation is close to adiabatic transformation. 
Our goal is to minimize $1-\Omega(s)$ and it is achieved by minimizing the action
\begin{equation}
\epsilon[\bm{\lambda}(s)]=\int_0^s\|[\partial_{s^\prime}P_n(\bm{\lambda};s^\prime),P_n(\bm{\lambda};s^\prime)]\|ds^\prime,
\end{equation}
where the norm $\|\cdot\|$ is the operator norm. 
We can rewrite this action as
\begin{equation}
\epsilon[\bm{\lambda}(s)]=\int_0^s\sqrt{2g_{ij}\dot{\lambda}^i\dot{\lambda}^j}ds^\prime,
\end{equation}
with the metric
\begin{equation}
g_{ij}=\mathrm{Re}\left[\sum_{m\ne n}\frac{\langle\Psi_n|\partial_i\mathcal{H}|\Psi_m\rangle\langle\Psi_m|\partial_j\mathcal{H}|\Psi_n\rangle}{(E_m-E_n)^2}\right], 
\end{equation}
which is known as the quantum geometric tensor. 
The Euler-Lagrange equation leads to the geodesic equation
\begin{equation}
\ddot{\lambda}^i+\Gamma^i_{jk}\dot{\lambda}^j\dot{\lambda}^k=0,
\label{Eq.geo.eq.multi}
\end{equation}
with
\begin{equation}
\Gamma^i_{jk}=\frac{1}{2}g^{il}(\partial_kg_{li}+\partial_jg_{lk}-\partial_lg_{jk}). 
\end{equation}
This geodesic equation gives an optimal schedule of parameters under a given initial condition (for the detailed derivation, see, Ref.~\cite{Rezakhani2010}).

%
%=====================================================================
%
\section{Results}
%
%--------------------------------------------------------------------------------------------------------------------------------------------
%
\subsection{\label{Sec.setup}Setups}
We rewrite the bosonic Josephson junction Hamiltonian (\ref{Eq.BJJ.ham}) as
\begin{equation}
\mathcal{H}_\mathrm{BJJ}=\mathrm{sgn}(\chi)\left(\frac{1}{N}J_z^2+\Lambda^{-1}J_x\right), 
\end{equation}
and set the rescaled time $\tau=|\chi|Nt$ with fixed $\chi$. 
Then, time dependence of the Hamiltonian only comes from $\Lambda^{-1}$. 
Note that this rescaled time differs from the normalized time in the previous section. 
We will simulate the rescaled Schr\"odinger equation
\begin{equation}
i\frac{\partial}{\partial\tau}|\Psi(\tau)\rangle=\mathcal{H}_\mathrm{BJJ}|\Psi(\tau)\rangle. 
\end{equation}
As the initial state, we prepare the coherent spin state along the $x$-axis
\begin{equation}
|\Psi(0)\rangle=2^{-N/2}\sum_{n=0}^N\sqrt{\binom{N}{n}}|J,J-n\rangle. 
\end{equation}
The parity of this state is $\Pi=+1$ and it is close to the ground state of the Hamiltonian with $|\Lambda^{-1}|>1$.

We determine time dependence of the parameter $\Lambda^{-1}$ according to the geodesic equation (\ref{Eq.geo.eq.multi}) for adiabaticity. 
For single parameter $\lambda=\Lambda^{-1}$, the geodesic equation (\ref{Eq.geo.eq.multi}) reduces to~\cite{Rezakhani2010}
\begin{equation}
2g\ddot{\lambda}+(\partial_{\lambda}g)\dot{\lambda}^2=0. 
\label{Eq.geo.eq.single.two}
\end{equation}
We can numerically solve this equation by giving the initial parameter $\lambda(0)$ and the time-derivative of the initial parameter $\dot{\lambda}(0)$. 
The following equality
\begin{equation}
\frac{d}{d\tau}(\sqrt{g}\dot{\lambda})=\frac{1}{2\sqrt{g}}[2g\ddot{\lambda}+(\partial_{\lambda}g)\dot{\lambda}^2]=0,
\end{equation}
holds, and thus we can also rewrite the geodesic equation (\ref{Eq.geo.eq.single.two}) as
\begin{equation}
\sqrt{g}\dot{\lambda}=C,
\label{Eq.QAB.simple}
\end{equation}
with a constant $C$. 
In this case, we can numerically solve this equation by giving the initial parameter $\lambda(0)$ and giving certain constant $C$ instead of $\dot{\lambda}(0)$. 
Note that the metric $g$ is simply given by
\begin{equation}
g=\sum_{m\neq n}\frac{|\langle\Psi_m|(\partial_\lambda\mathcal{H}_\mathrm{BJJ})|\Psi_n\rangle|^2}{(E_m-E_n)^2}. 
\label{Eq.metric.BJJ}
\end{equation}
Because parameter-derivative of the Hamiltonian $(\partial_\lambda\mathcal{H}_\mathrm{BJJ})$ and the parity operator $\Pi$ are commutative, i.e., 
\begin{equation}
[(\partial_\lambda\mathcal{H}_\mathrm{BJJ}),\Pi]=0, 
\end{equation}
we can reduce summation in the metric (\ref{Eq.metric.BJJ}) so that the eigenstate $|\Psi_m\rangle$ has the identical parity to $|\Psi_n\rangle$. 
This reduction is quite important for the present generation schemes because we can avoid numerical divergence associated with gap closing and with the degeneracies of the eigenstates. 

%
%----------------------------------------------------------------------------------------------------------------------------------------------
%
\subsection{\label{Sec.cat}Generation of cat states}
\begin{figure}
\includegraphics[width=8.5cm]{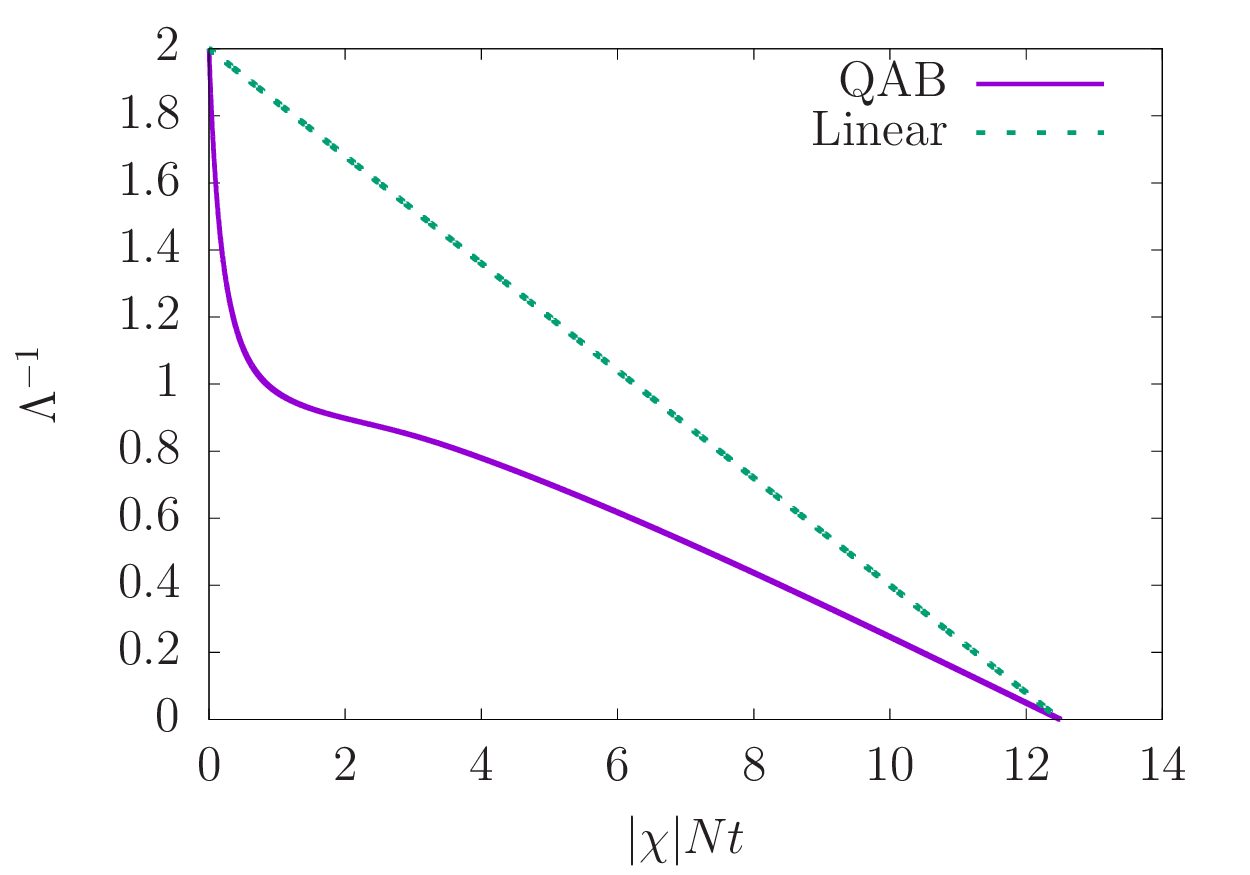}
\caption{\label{Fig.schedule-N100C0_5} An example of schedules of $\Lambda^{-1}$ with respect to the rescaled time $|\chi|Nt$. The purple solid curve represents the schedule based on quantum adiabatic brachistochrone for $C=-0.5$ and the green dotted line represents the linear schedule with the same generation time. Here $N=100$. }
\end{figure}

\begin{figure}
\includegraphics[width=8.5cm]{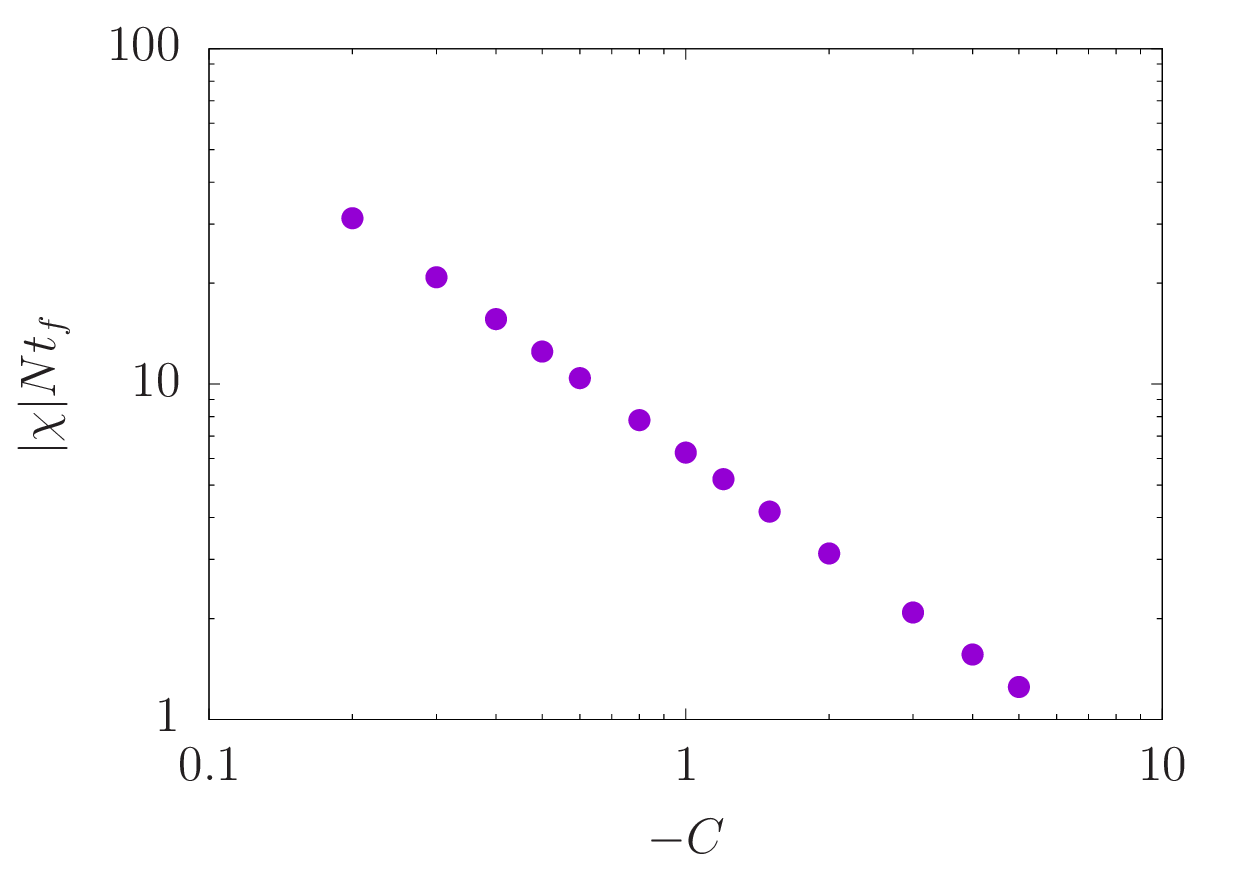}
\caption{\label{Fig.tfC-N100} The constant $C$ dependence of the rescaled generation time $|\chi|Nt_f$. Here $N=100$. }
\end{figure}

First we consider negative nonlinearity $\chi<0$ to create a cat-like state. 
We change the parameter $\Lambda^{-1}$ from $2$ to $0$. 
An example of schedules optimized by quantum adiabatic brachistochrone is shown in Fig.~\ref{Fig.schedule-N100C0_5} with the corresponding linear schedule, $\lambda(\tau)=(\tau/\tau_f)\lambda(\tau_f)+(1-\tau/\tau_f)\lambda(0)$, where $\tau_f=|\chi|Nt_f$ is the generation time. 
Here we set $\lambda(0)=2$, $\lambda(\tau_f)=0$, and $C=-0.5$. 
Quantum adiabatic brachistochrone suggests fast decrease before approaching the critical point, $\Lambda^{-1}\gtrsim\Lambda_c^{-1}=1$ (in the disordered phase), and slow decrease after passing there (above the critical point and in the ordered phase). 
We can observe similar schedules with different generation time $t_f$ when we set other values of the constant $C$. 
The relationship between the generation time $t_f$ and the constant $C$ is shown in Fig.~\ref{Fig.tfC-N100}. 
It clearly indicates $t_f\propto C^{-1}$. 
Indeed, from Eq.~(\ref{Eq.QAB.simple}), we can obtain
\begin{equation}
\tau=C^{-1}\int_{\lambda(0)}^{\lambda(\tau)}\sqrt{g}d\lambda. 
\end{equation}
Because $g$ is a function of $\lambda$, the generation time $t_f$ is inversely proportional to the constant $C$.

\begin{figure}
\includegraphics[width=8.5cm]{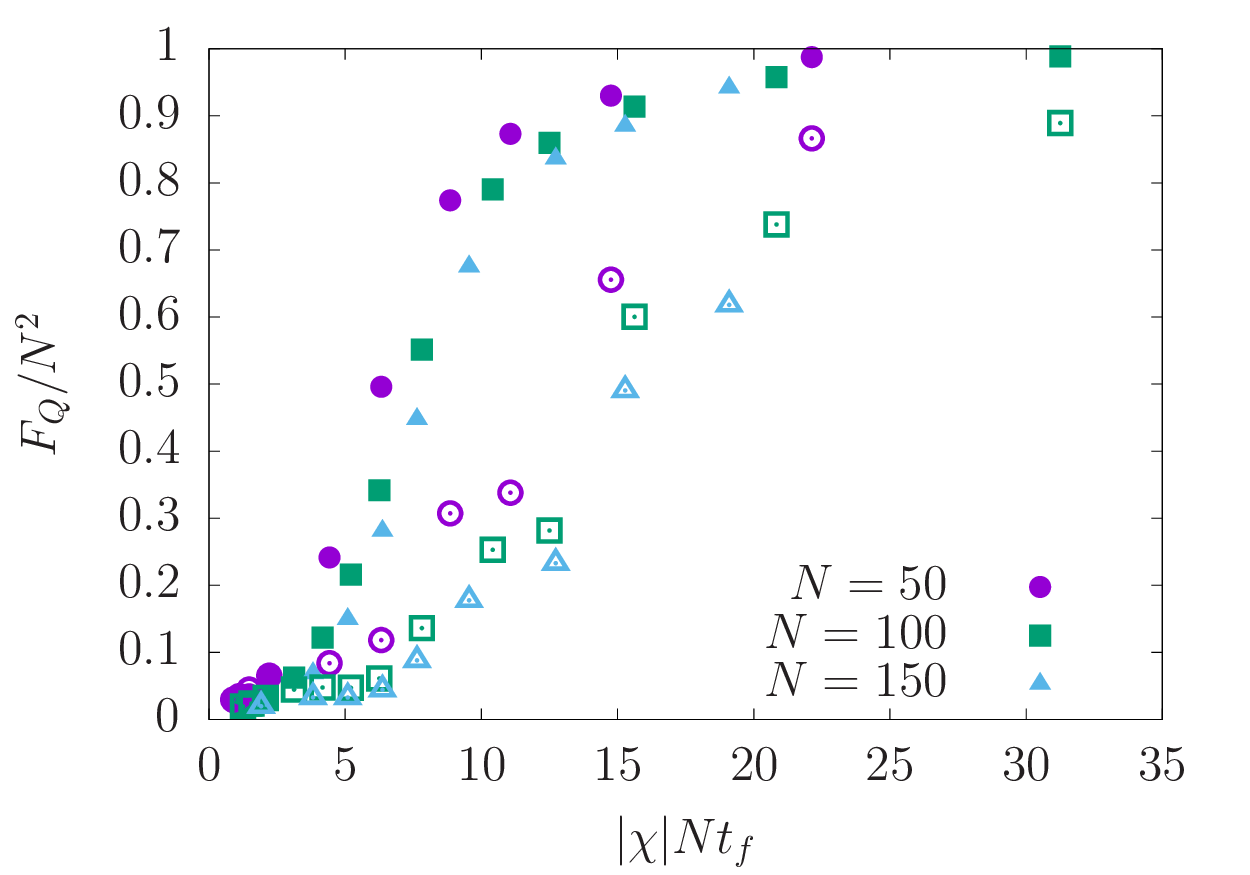}
\caption{\label{Fig.qFishertf-N100} The rescaled quantum Fisher information $F_Q/N^2$ with respect to rescaled generation time $|\chi|Nt_f$. The filled symbols represent quantum adiabatic brachistochrone and the open symbols represent the linear schedule. Here $N=50$ (purple circles), $100$ (green squares), and $150$ (cyan triangles). }
\end{figure}

\begin{figure}
\includegraphics[width=8.5cm]{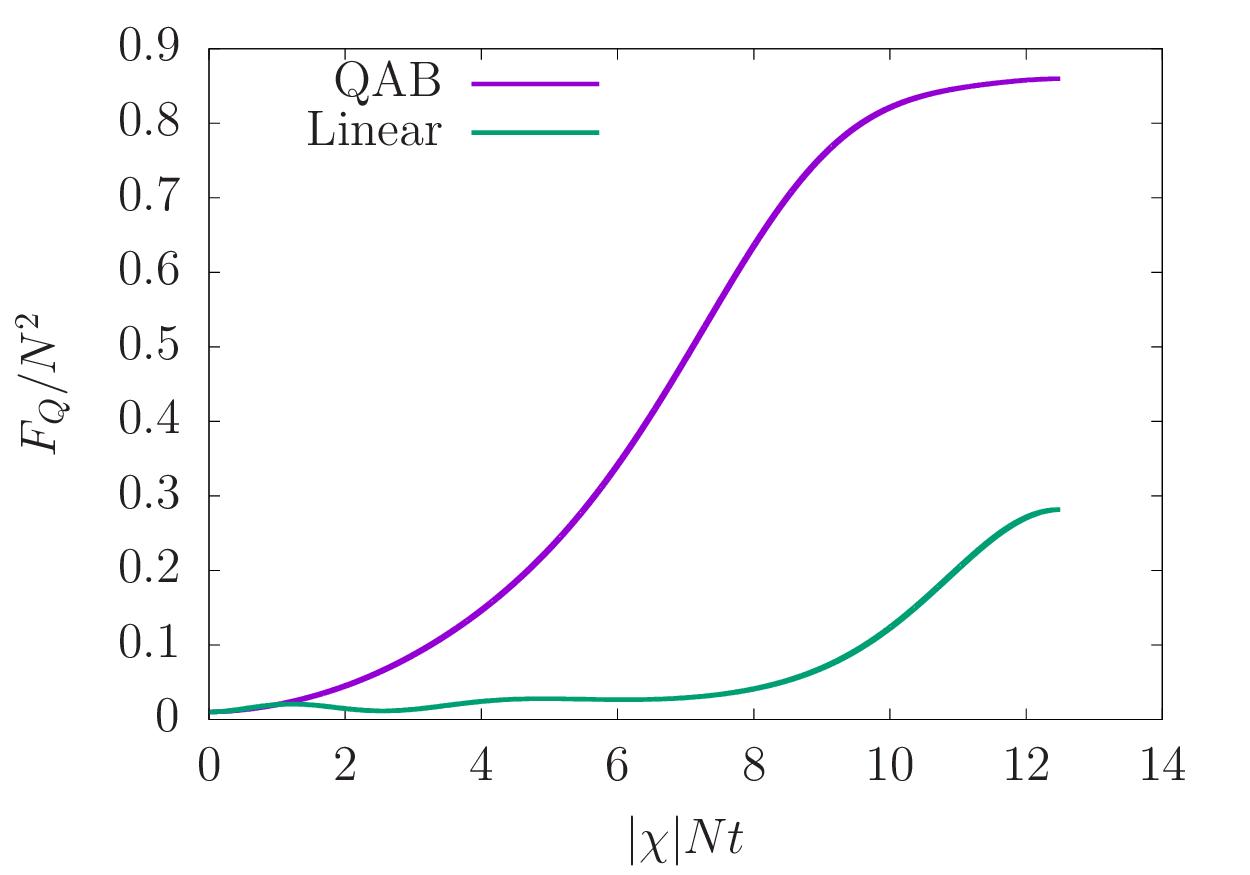}
\caption{\label{Fig.qfishertevo} Growth of quantum Fisher information with respect to the rescaled time $|\chi|Nt$. The purple curve represents quantum adiabatic brachistochrone and the green curve represents the linear schedule. Here $C=-0.5$ and $N=100$. }
\end{figure}

Now we study generation of a cat-like state according to the schedules obtained by quantum adiabatic brachistochrone (Fig.~\ref{Fig.schedule-N100C0_5}). 
Here we calculate the quantum Fisher information $F_Q$, which is a measure of macroscopicity of entanglement and related to uncertainty of estimation as $\Delta\theta=1/\sqrt{F_Q}$ for an unknown parameter $\theta$, with various values of the constant $C$. 
We set the interferometric axis to $J_z$, i.e., the quantum Fisher information is given by
\begin{equation}
F_Q[|\Psi(\tau_f)\rangle,J_z]=4(\Delta J_z)^2,
\label{Eq.qFisher.Jz}
\end{equation}
where $(\Delta J_z)^2$ is the variance of $J_z$ with the state $|\Psi(\tau_f)\rangle$. 
Here, from Eq.~(\ref{Eq.qFisher.Jz}), the maximum value of the quantum Fisher information with $J_z$ is $N^2$. 
It is known that with a local operator, such as $J_z$, a given state is at least entangled if the quantum Fisher information is larger than $N$ and a given state is macroscopically entangled if the quantum Fisher information scales as $N^2$ (see, e.g., Ref.~\cite{Pezze2018}). 
In the present generation scheme, the large quantum Fisher information ensures the low excess energy, i.e., the generated state is close to a cat state, and potential usefulness in interferometry using the parity measurement~\cite{Hatomura2019}. 
Here we plot the rescaled quantum Fisher information $F_Q/N^2$ with respect to the rescaled generation time $|\chi|Nt_f$ in Fig.~\ref{Fig.qFishertf-N100}, where the filled symbols represent the results by quantum adiabatic brachistochrone and the open symbols represent the results by the linear schedule. 
Clearly quantum adiabatic brachistochrone improves adiabaticity. 
Note that quantum Fisher information monotonically increases during generation with quantum adiabatic brachstochrone as seen in Fig.~\ref{Fig.qfishertevo}, where the rescaled quantum Fisher information $F_Q/N^2$ is plotted with respect to the rescaled time $|\chi|Nt$. 

\begin{figure}
\includegraphics[width=8.5cm]{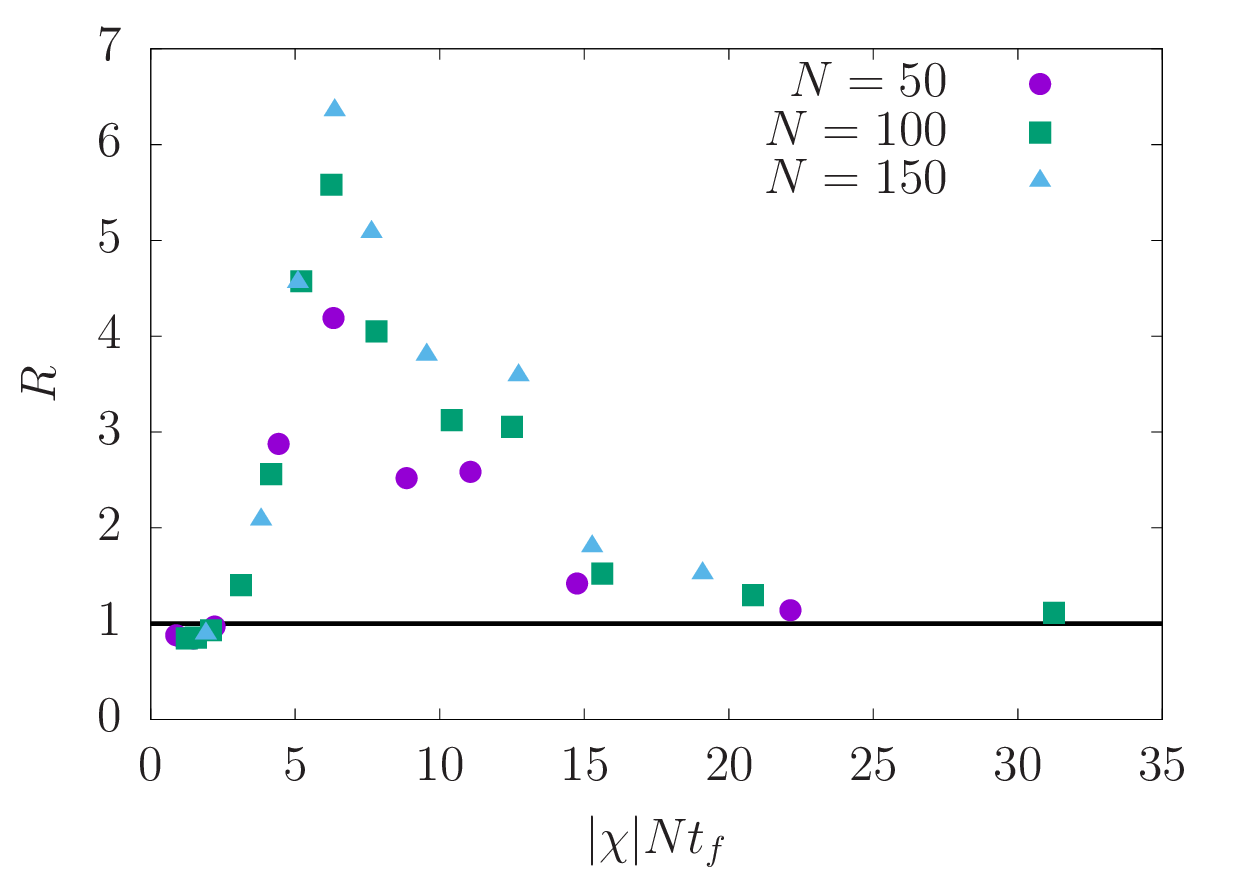}
\caption{\label{Fig.qFisherN} The rate of improvement in quantum Fisher information $R=F_Q^\mathrm{QAB}/F_Q^\mathrm{Linear}$ with respect to the rescaled generation time $\tau_f=|\chi|Nt_f$. Here $N=50$ (purple circles), $100$ (green squares), and $150$ (cyan triangles). We also plot supplemental black line $R=F_Q^\mathrm{QAB}/F_Q^\mathrm{Linear}=1$. }
\end{figure}

It is of interest if the present results can scale with system size or not. 
In Fig.~\ref{Fig.qFishertf-N100}, we find that the rescaled quantum Fisher information $F_Q/N^2$ against the rescaled generation time $|\chi|Nt_f$ does not make big difference except for small finite-size corrections when we change the system size $N$ from $50$ to $150$. 
Here we further study how much quantum adiabatic brachistochrone improves the quantum Fisher information compared with that of the linear schedule. 
We plot the improving rate $R=F_Q^\mathrm{QAB}/F_Q^\mathrm{Linear}$ in Fig.~\ref{Fig.qFisherN}, where $F_Q^\mathrm{QAB}$ ($F_Q^\mathrm{Linear}$) is the quantum Fisher information of the generated state by quantum adiabatic brachistochrone (the linear schedule). 
The improving rate $R$ also shows similar behavior when we change the system size $N$. 
The important point is that quantum adiabatic brachistochrone improves quantum Fisher information more than double in the fast generation time regime $4\lesssim|\chi|Nt_f\lesssim14$, where the large quantum Fisher information can be obtained by quantum adiabatic brachistochrone. 
It is also important that the rate of improvement becomes large when we increase the system size $N$.

\begin{figure}
\includegraphics[width=8.5cm]{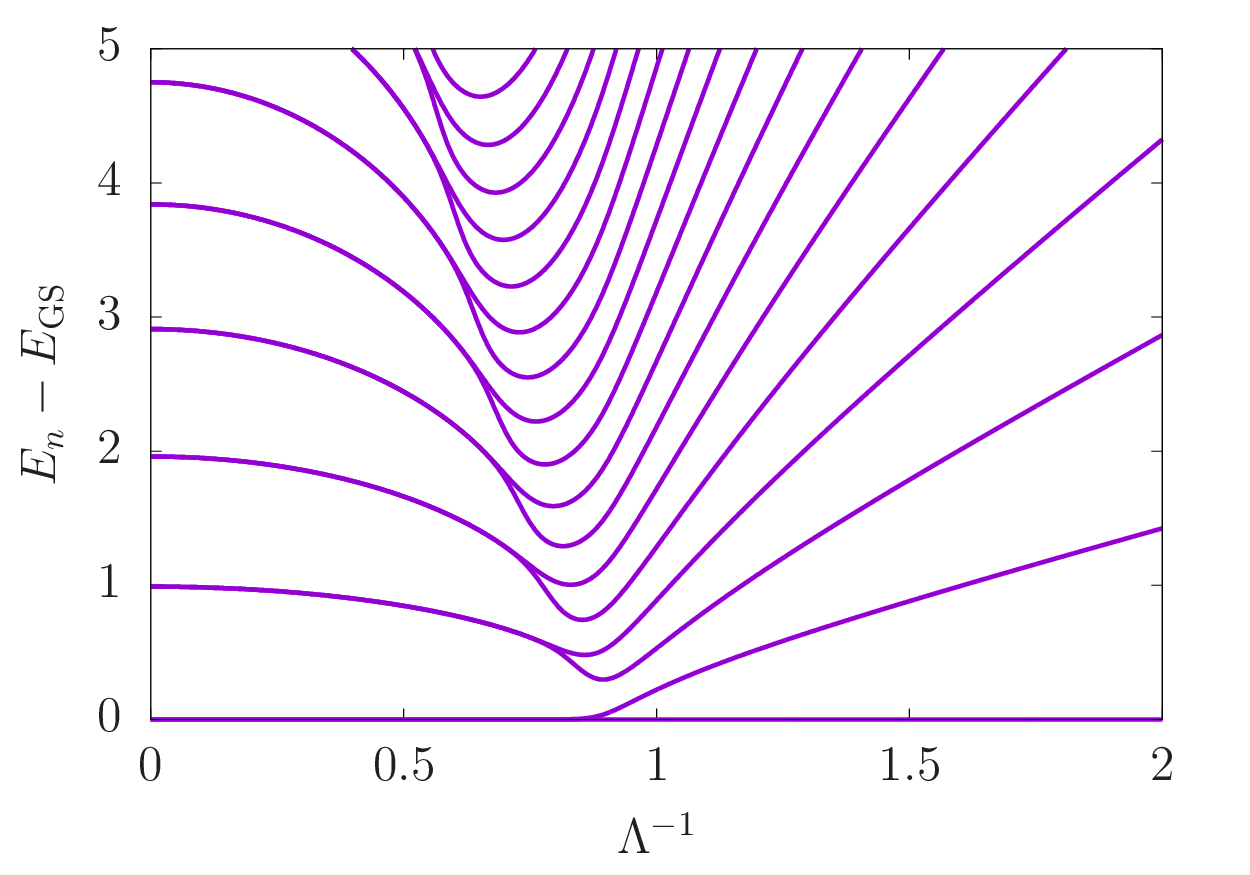}
\caption{\label{Fig.BJJene-N100} The energy spectrum $\{E_n-E_0\}$ with respect to the parameter $\Lambda^{-1}$. Here $N=100$. }
\end{figure}

\begin{figure}
\includegraphics[width=8.5cm]{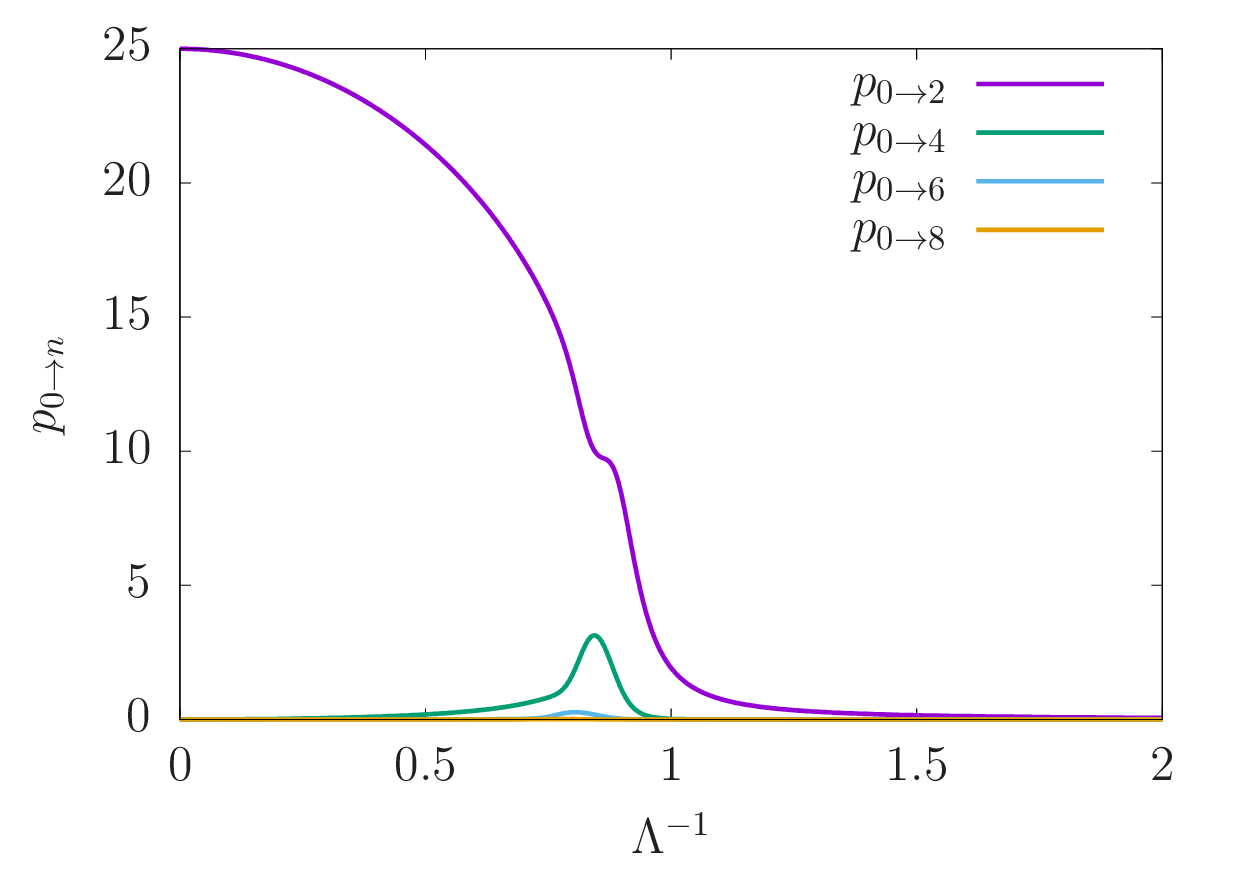}
\caption{\label{Fig.BJJmat-N100} A part of the transition matrices from the ground state to excited states as functions of $\Lambda^{-1}$. Here $N=100$. }
\end{figure}

We now discuss why the schedules obtained by quantum adiabatic brachistochrone take the form of Fig.~\ref{Fig.schedule-N100C0_5} and why they can improve adiabaticity. 
We plot a part of the energy spectra in Fig.~\ref{Fig.BJJene-N100}. 
Clearly it shows small energy gaps around the critical point $\Lambda_c^{-1}=1$ (although there are some finite-size corrections), and thus we have to slowly change the parameter around there. 
We can see increase of the energy differences when the parameter becomes small, but quantum adiabatic brachistochrone still suggests to change the parameter slowly. 
We can confirm this reason by calculating the transition matrices
\begin{equation}
p_{0\to n}=|\langle\Psi_n(\tau)|(\partial_\lambda\mathcal{H}_\mathrm{BJJ})|\Psi_0(\tau)\rangle|^2. 
\end{equation}
Here we plot some of main contributions in Fig.~\ref{Fig.BJJmat-N100}. 
We find that the transition matrix from the ground state to the second excited state $p_{0\to2}$ becomes large when the parameter $\Lambda^{-1}$ decreases in contrast to decrease of other elements. 
This is why quantum adiabatic brachistochrone suggests slow decrease even though energy differences become large. 
Note that the transition matrices from the ground state to odd numbered excited states vanish due to the parity conservation.

%
%--------------------------------------------------------------------------------------------------------------------------------------------------------------------------------------------------------
%
\subsection{\label{Sec.squeezed}Generation of spin squeezed states}

\begin{figure}
\includegraphics[width=8.5cm]{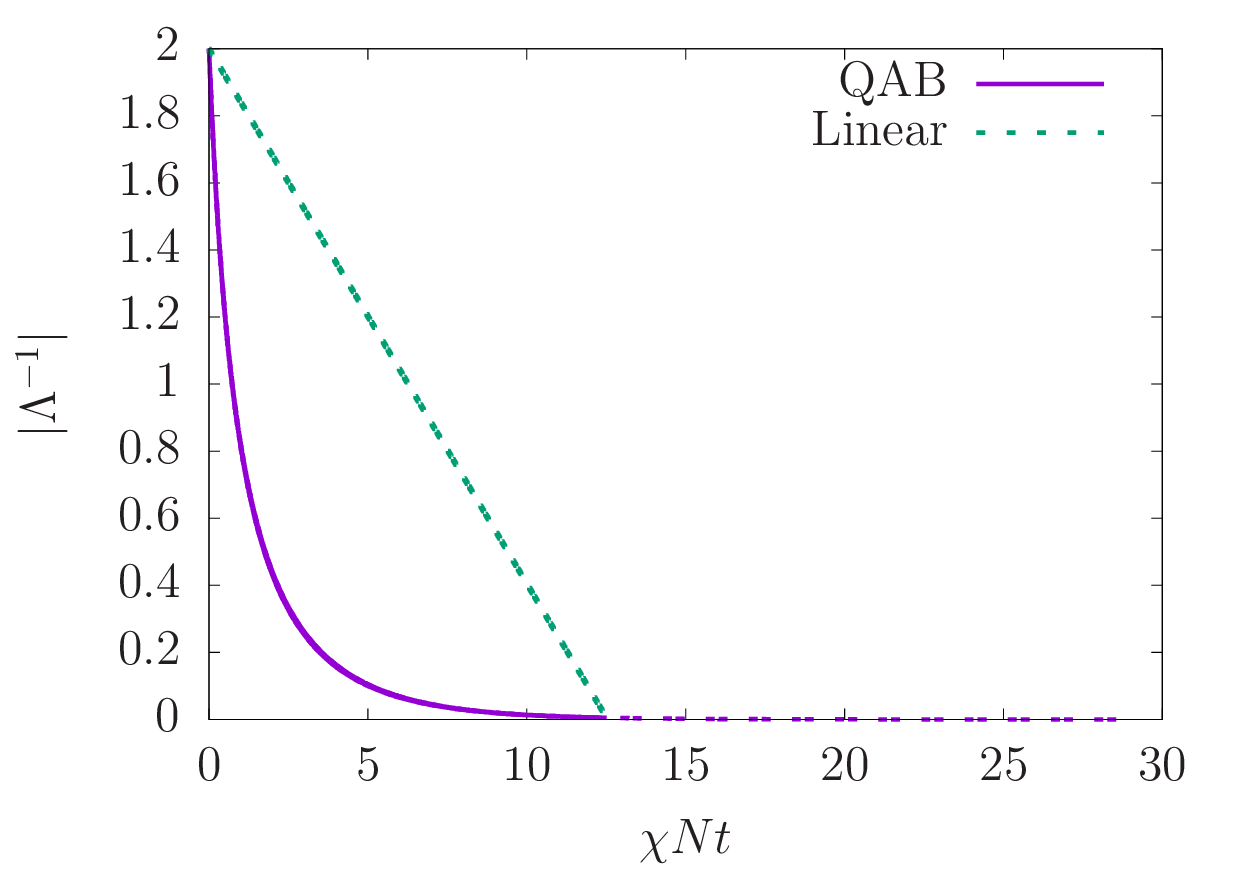}
\caption{\label{Fig.squeeze-schedule} An example of schedules of $|\Lambda^{-1}|$ with respect to the rescaled time $\chi Nt$. The purple solid curve represents the schedule based on quantum adiabatic brachistochrone during $|\Lambda^{-1}|:2\to0.005$ and the purple dotted curve, which is extended from the solid curve, is that during $|\Lambda^{-1}|:0.005\to0$, for $C=-0.07$. The green dotted line is the corresponding linear schedule to the purple solid curve. Here $N=100$. }
\end{figure}

\begin{figure}
\includegraphics[width=8.5cm]{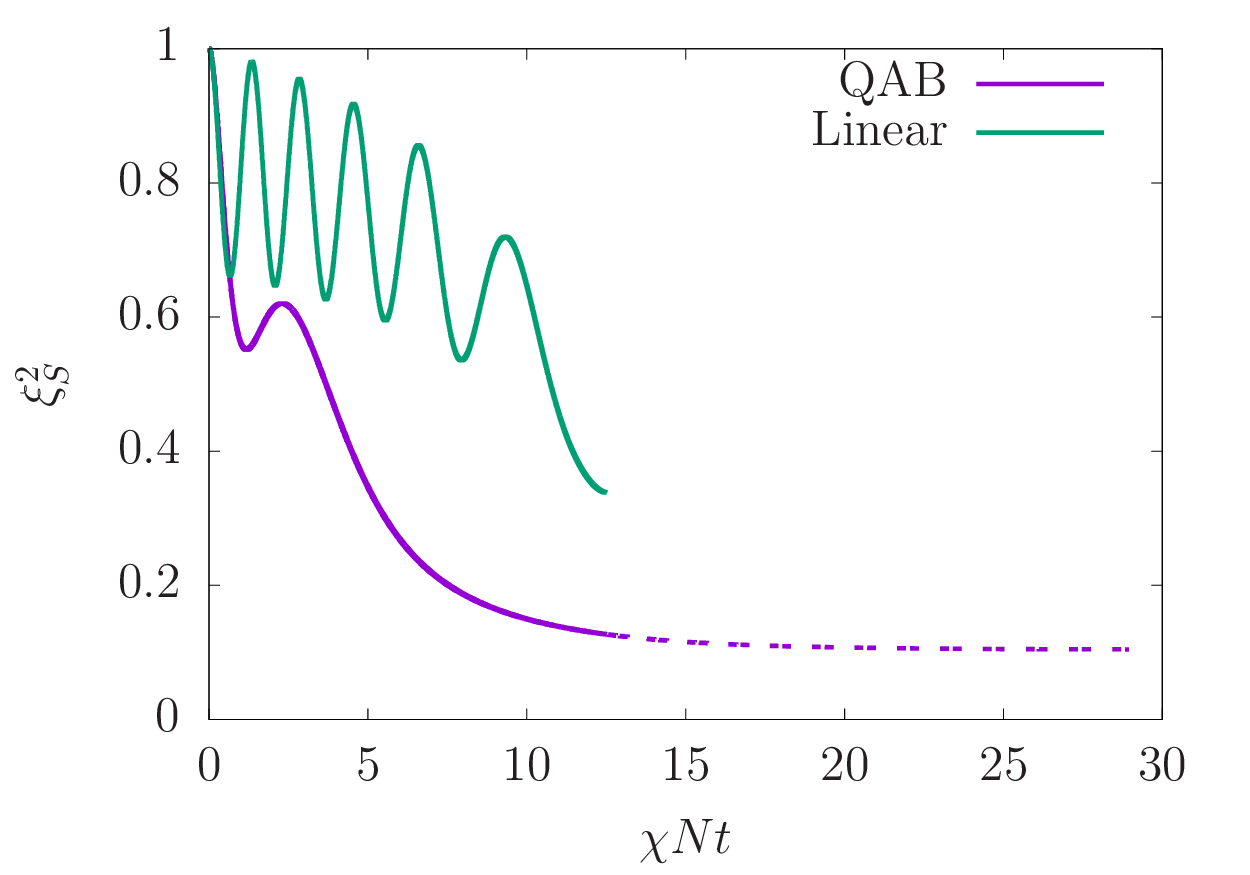}
\caption{\label{Fig.Wparameter.tevo} Decrease of the spin squeezing parameter with respect to the rescaled time $\chi Nt$. The purple solid curve represents quantum adiabatic brachistochrone with $|\Lambda^{-1}|:2\to0.005$ and the purple dotted curve extended from the solid curve is that with $|\Lambda^{-1}|:0.005\to0$. The green solid curve represents the corresponding linear schedule with $|\Lambda^{-1}|:2\to0.005$. Here $C=-0.07$ and $N=100$. }
\end{figure}

Next, we consider positive nonlinearity $\chi>0$ to create a spin squeezed state. 
An example of schedules optimized by quantum adiabatic brachistochrone is shown in Fig.~\ref{Fig.squeeze-schedule}. 
In contrast to the case of negative nonlinearity $\chi<0$, quantum adiabatic brachistochrone just suggests fast decrease at first and extremely slow decrease around $|\Lambda^{-1}|\approx0$. 
However, this final slow process does not contribute so much to spin squeezing. 
In order to show it, we consider two cases, $|\Lambda^{-1}|:2\to0$ and $|\Lambda^{-1}|:2\to0.005$, with $C=-0.07$. 
The operation time of the case $|\Lambda^{-1}|:2\to0$ is about $\chi Nt_f=29.0$, while that of the case $|\Lambda^{-1}|:2\to0.005$ is about $\chi Nt_f=12.5$. 
To decrease parameter $|\Lambda^{-1}|$ from $0.005$ to $0$, we need an additional generation time $\chi Nt=16.5$, which is longer than time to decrease $|\Lambda^{-1}|$ from $2$ to $0.005$. 
Nevertheless, the spin squeezing parameter
\begin{equation}
\xi_S^2=\frac{N(\Delta J_z)^2}{\langle J_x\rangle^2}, 
\end{equation}
which is related to uncertainty of estimation as $\Delta\theta=\xi_S/\sqrt{N}$ for an unknown parameter $\theta$ and thus a given state is metrologically useful if $\xi_S<1$, does not change so much during the process $|\Lambda^{-1}|:0.005\to0$ as shown in Fig.~\ref{Fig.Wparameter.tevo}. 
Therefore, hereafter we only consider the process $|\Lambda^{-1}|:2\to0.005$. 
Note that in contrast to the case of the quantum Fisher information for negative nonlinearity $\chi<0$, the spin squeezing parameter does not monotonically decrease during generation by quantum adiabatic brachistochrone. 
This is because both of $(\Delta J_z)$ and $\langle J_x\rangle$ decrease and the spin squeezing parameter depends on the rate of these quantities.

\begin{figure}
\includegraphics[width=8.5cm]{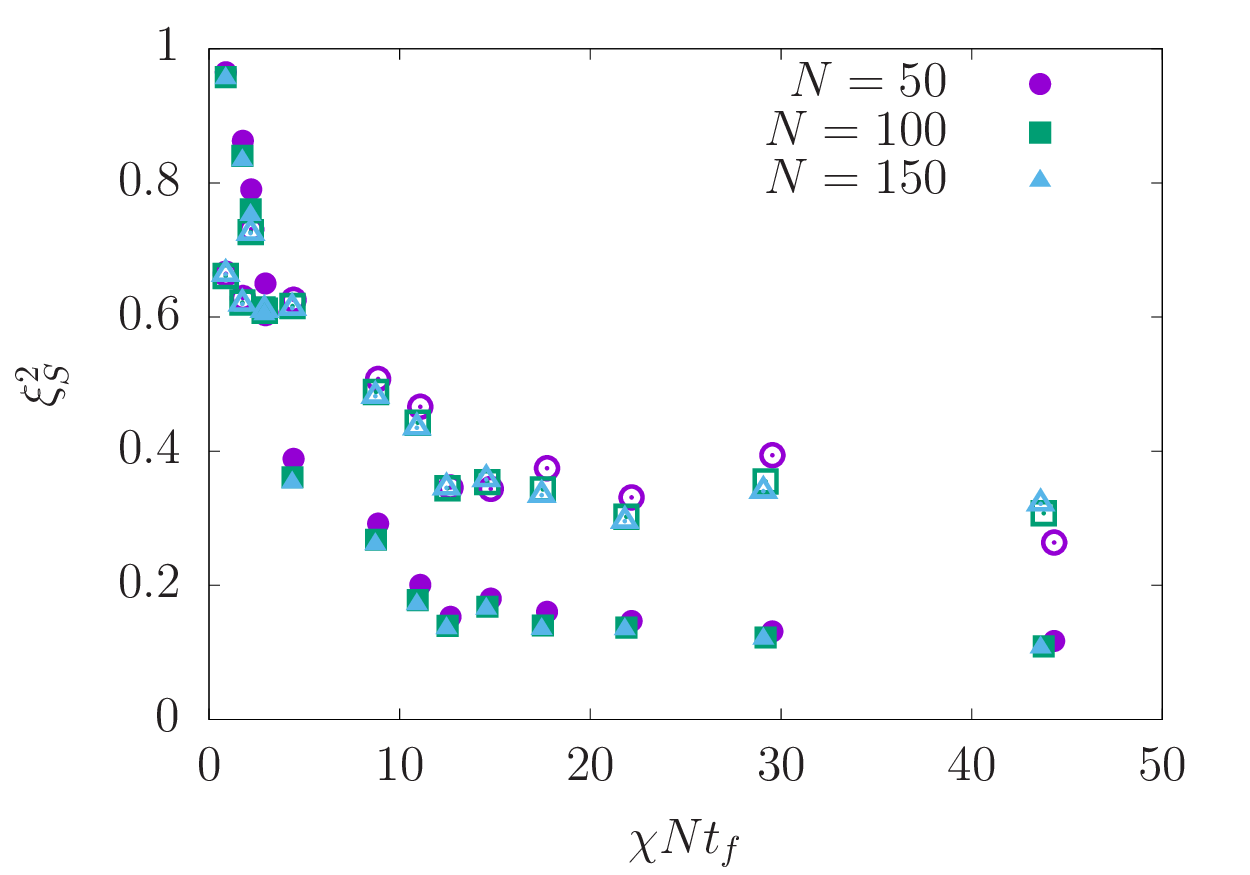}
\caption{\label{Fig.Wparameter} The square of the spin squeezing parameter $\xi_S^2$ with respect to the rescaled generation time $\chi Nt_f$. The filled symbols represent quantum adiabatic brachistochrone and the open symbols represent the linear schedule. Here $N=50$ (purple circles), $100$ (green squares), and $150$ (cyan triangles). }
\end{figure}

\begin{figure}
\includegraphics[width=8.5cm]{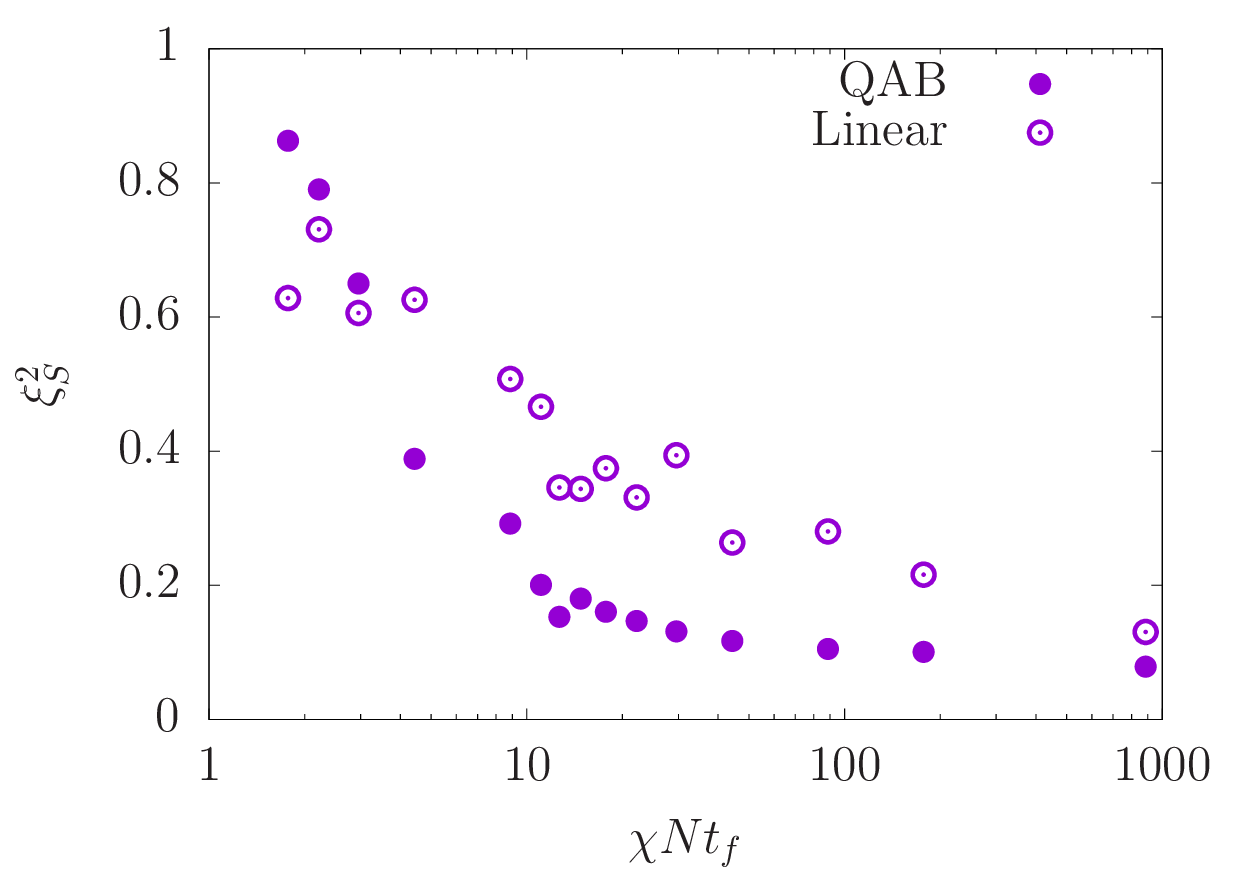}
\caption{\label{Fig.Wparameter2} The longer time version of Fig.~\ref{Fig.Wparameter} for $N=50$. The filled symbols represent quantum adiabatic brachistochrone and the open symbols represent the linear schedule. }
\end{figure}

We calculate the spin squeezed parameter with various values of the constant $C$ and depict it in Fig.~\ref{Fig.Wparameter}. 
As we can see, quantum adiabatic brachistochrone decreases the spin squeezed parameter, i.e., increases metrological usefulness, except for an ultrafast regime $\chi Nt_f\lesssim3$. 
In this ultrafast regime, the spin squeezed parameter is not so small, and thus we are not so interested in there. 
Therefore, we can generally say that quantum adiabatic brachistochrone improves generation of a spin squeezed state. 
In contrast to the case of negative nonlinearity $\chi<0$, system size dependence is quite small. 
For small number of particles, $N=50$, we calculate the square of the spin squeezing parameter over the long time regime and plot it in Fig.~\ref{Fig.Wparameter2}. 
Notably, the spin squeezing parameter of a spin squeezed state generated via quantum adiabatic brachistochrone with $\chi Nt_f\sim20$ is approximately equal to that generated by the linear schedule with $\chi Nt_f\sim900$. 
This result shows significance to use quantum adiabatic brachistochrone for generation of a spin squeezed state.

\begin{figure}
\includegraphics[width=8.5cm]{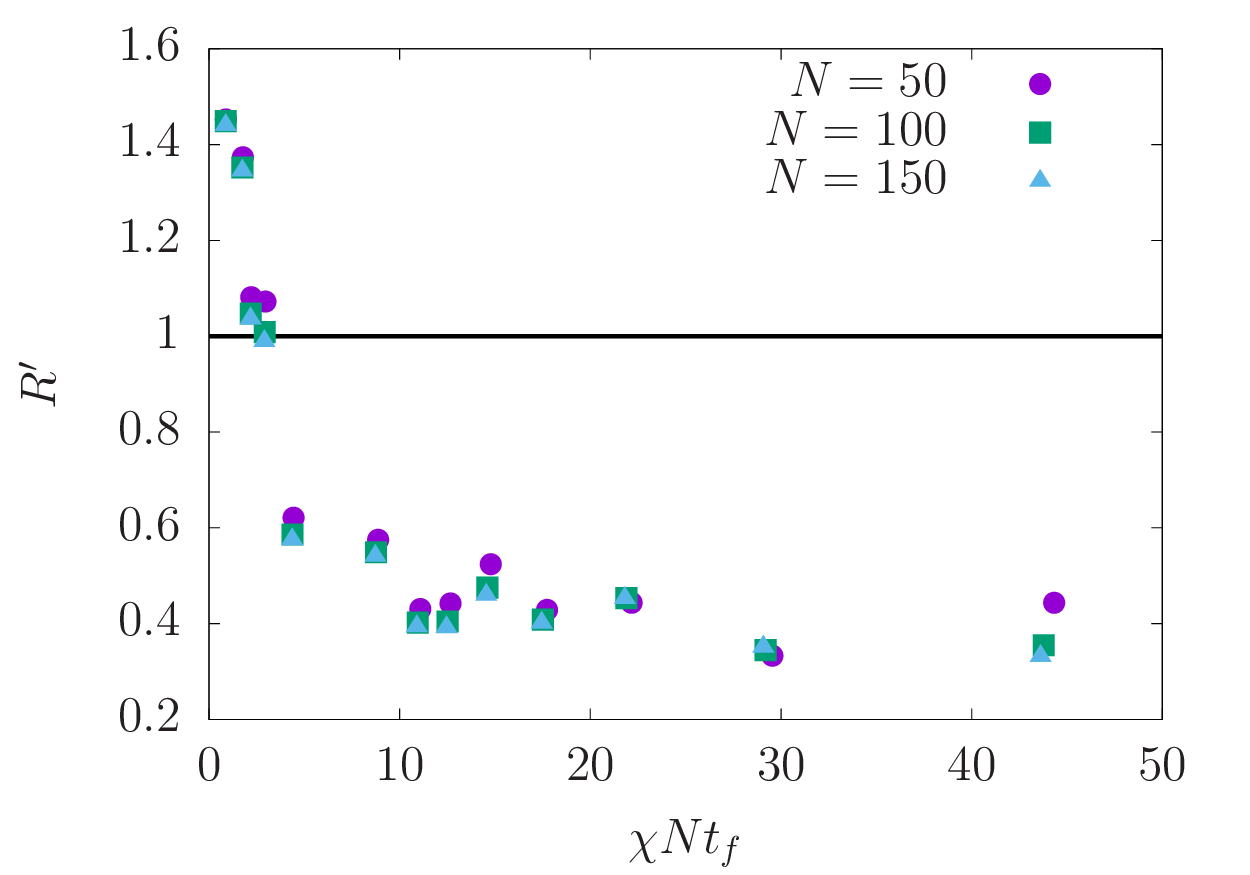}
\caption{\label{Fig.Wrate} The rate of improvement in the spin squeezing parameter $R^\prime=(\xi_S^\mathrm{QAB}/\xi_S^\mathrm{Linear})^2$ with respect to the generation time $\chi Nt_f$. Here $N=50$ (purple circles), $100$ (green squares), and $150$ (cyan triangles). }
\end{figure}

We also calculate the improving rate of the spin squeezing parameter $R^\prime=(\xi_S^\mathrm{QAB}/\xi_S^\mathrm{Linear})^2$ and plot it in Fig.~\ref{Fig.Wrate}. 
For a wide generation time region, a spin squeezed state generated via quantum adiabatic brachistochrone has more than double metrological usefulness compared with that generated by the linear schedule.

\begin{figure}
\includegraphics[width=8.5cm]{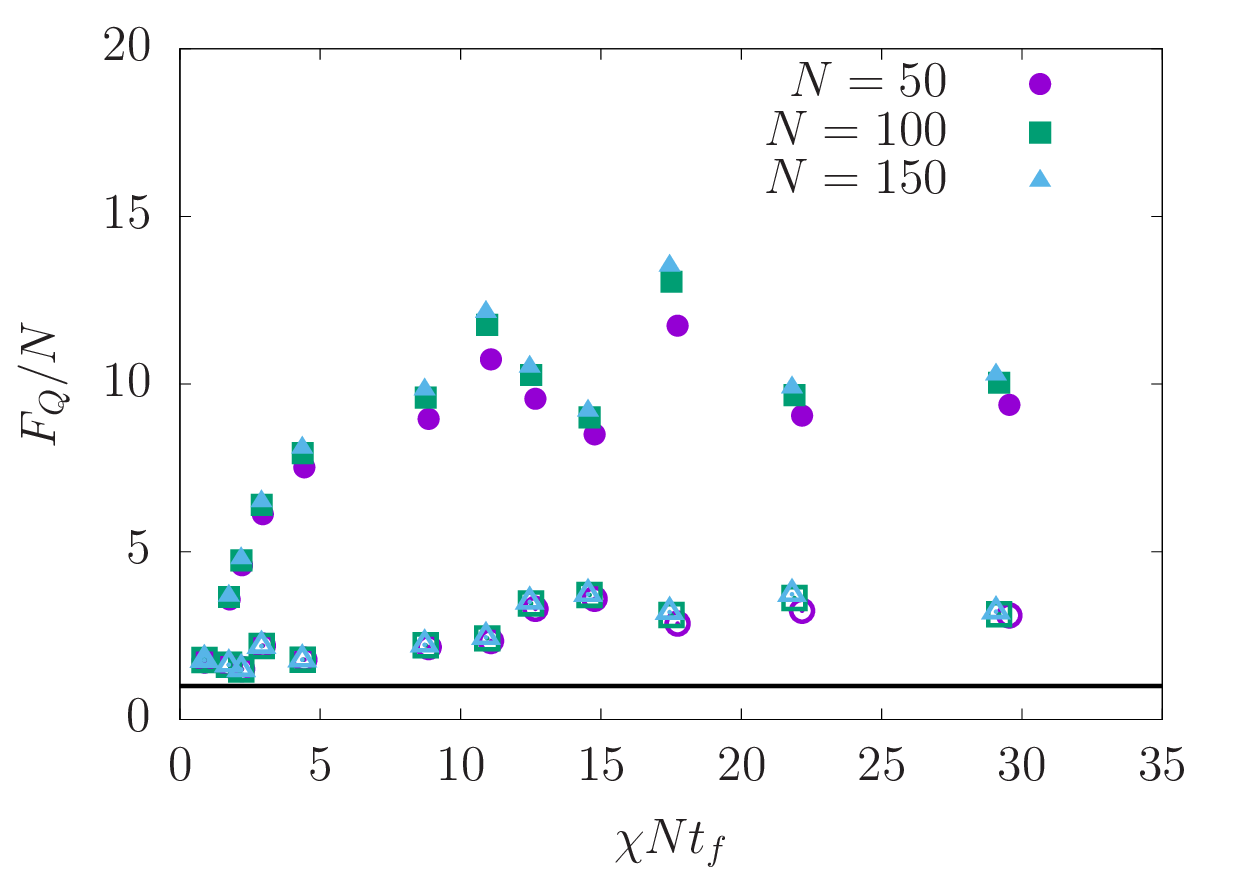}
\caption{\label{Fig.squeeze_qFisher}The quantum Fisher information scaled by $N$, $F_Q/N$, with respect to rescaled generation time $\chi Nt_f$. The filled symbols represent quantum adiabatic brachistochrone and the open symbols represent the linear schedule. Here $N=50$ (purple circles), $100$ (green squares), and $150$ (cyan triangles). Generated states are at least entangled and metrologically useful if $F_Q/N$ surpasses the black solid line $F_Q/N=1$. }
\end{figure}

Finally, we also calculate the quantum Fisher information of the generated spin squeezed states to compare with the generated cat-like states. 
It should be noted that the quantum Fisher information of a given state is larger than $N$, i.e., it is metrologically useful, if the spin squeezing parameter of the state is less than $1$ because the following inequality $F_Q\ge N/\xi_S^2$ holds~\cite{Pezze2018}. 
Here we plot the quantum Fisher information scaled by $N$, $F_Q/N$, in Fig.~\ref{Fig.squeeze_qFisher}. 
It clearly shows that the generated spin squeezed states are metrologically useful and quantum adiabatic brachistochrone also improves the quantum Fisher information of spin squeezed states. 
However, it should be noted that the size of the quantum Fisher information of spin squeezed states is not so large compared with that of cat-like states. 
Indeed, for a fixed scaled generation time $|\chi|Nt_f$, the quantum Fisher information of the generated spin squeezed states scales as $F_Q\propto N$ with $F_Q>N$, whereas that of the generated cat-like states scales as $F_Q\propto N^2$. 
Moreover, the maximum size of the quantum Fisher information of a spin squeezed state as the ground state of the bosonic Josephson junction Hamiltonian is $F_Q=N^2/2+N$~\cite{Pezze2018}, while that of a cat state is $F_Q=N^2$. 

%
%======================================================================
%
\section{\label{Sec.summary}Summary}

In this paper, we considered quantum adiabatic brachistochrone in a bosonic Josephson junction to find time-optimal schedules to create a cat-like state and a spin squeezed state. 
In contrast to counterdiabatic driving in a bosonic Josephson junction, which requires the two-axis countertwisting interaction~\cite{Hatomura2018a}, quantum adiabatic brachistochrone does not require any additional term. 
Moreover, we found that the optimized schedules are smooth and monotonically decreasing curves, which is rather realistic to implement in experiments compared with schedules designed by conventional optimal control theory~\cite{Grond2009a,Pichler2016,Lapert2012}. 
Note that, for creation of a cat-like state, the parity conservation plays an important role to avoid divergence of the quantum geometric tensor. 
It is also advantageous that generated states are automatically trapped by classical fixed points after generation. 
In both generation of a cat-like state and a spin squeezed state, we observed improvement of adiabaticity and increase of metrological usefulness compared with corresponding linear schedules.

Finally we discuss possibility to realize the present schemes in experiments. 
Our numerical simulation was studied by using the scaled time $|\chi|Nt$, and thus we first consider accessible generation time from the viewpoint of nonlinearity. 
In Refs.~\cite{Gross2010,Zibold2010}, a Bose-Einstein condensate of $^{87}\mathrm{Rb}$ shows $\chi=2\pi\times0.063~\mathrm{Hz}$ for $N=200\sim450$ atoms. 
Therefore, the prefactor of the rescaled time can achieve $\chi N\approx178~\mathrm{s}^{-1}$. 
Similarly, in Ref.~\cite{Strobel2014}, it shows $\chi N\approx188~\mathrm{s}^{-1}$. 
For a time generating an intermediate scale cat-like state (about 80 - 90 \% size of the largest cat state) or a well squeezed state ($\xi_S^2\le0.2$), $|\chi|Nt_f\approx10$, these values requires generation time $t_f\approx50~\mathrm{ms}$. 
Moreover, in Ref.~\cite{Riedel2010}, it shows $\chi=0.49~\mathrm{s}^{-1}$ with $N\approx1250$ by using the atom-chip-based technique and it achieves $\chi N\approx613~\mathrm{s}^{-1}$. 
For this parameter, the generation time is $t_f\approx16~\mathrm{ms}$. 
Note that larger number of atoms leads to faster decay of entanglement. 
Therefore, one might be rather interested in generation with several hundred atoms. 
According to Ref.~\cite{Strobel2014}, the nonlinearity $\chi$ behaves as $\chi\propto1/\sqrt{N}$, and thus it could be possible to achieve $\chi N\approx368~\mathrm{s}^{-1}$ for $N=450$ by using the atom-chip-based technique. 
It offers the generation time $t_f\approx27~\mathrm{ms}$. 
Entanglement generation experiments are typically performed about 10 - 50 ms~\cite{Esteve2008,Gross2010,Riedel2010,Maussang2010,Ockeloen2013,Strobel2014,Muessel2014,Muessel2015,Schmied2016} and generated states could be detectable even if about 10 \% of atoms are lost~\cite{Hatomura2019}, and thus we expect that generation time of the present scheme is realistic. 
Note that we have to simultaneously induce both of nonlinearity and the Rabi coupling as demonstrated in Ref.~\cite{Zibold2010}, and thus not all above parameters are available. It should be also noted that negative nonlinearity does not directly realize in the above experimental setups. However, we can realize the present scheme with negative nonlinearity by considering the highest energy eigenstate with positive nonlinearity as experimentally studied in Ref.~\cite{Zibold2010} and as theoretically studied in Ref.~\cite{Yukawa2018} because the ground state with negative nonlinearity and the highest energy eigenstate with positive nonlinearity are mathematically equivalent. It can be done by preparing the coherent spin state along the $(-x)$-axis instead of the $x$-axis as the initial state. 
Totally, we believe that an intermediate scale cat-like state and a well squeezed state will be generated by using our scheme in the near future.

% Specify following sections are appendices. Use \appendix* if there
% only one appendix.
%\appendix

% If you have acknowledgments, this puts in the proper section head.
%\begin{acknowledgments}
% put your acknowledgments here.
%\end{acknowledgments}

% Create the reference section using BibTeX:
\bibliography{BECcatQABbib}

\end{document}